\documentclass[%
 jmp,
 amsmath,amssymb,amsart
preprint,%
]{revtex4-1}

\usepackage{graphicx}
\usepackage{dcolumn}
\usepackage{bm}

\usepackage[utf8]{inputenc}
\usepackage[T1]{fontenc}
\usepackage{mathptmx}
\usepackage{etoolbox}
\usepackage{color}

\makeatletter
\def\@email#1#2{%
 \endgroup
 \patchcmd{\titleblock@produce}
  {\frontmatter@RRAPformat}
  {\frontmatter@RRAPformat{\produce@RRAP{*#1\href{mailto:#2}{#2}}}\frontmatter@RRAPformat}
  {}{}
}%
\makeatother

\newcommand\xx{{\text{\sf X}}}
\newcommand\lan{\langle}
\newcommand\ran{\rangle}
\newcommand\tr{{\text{\rm Tr}}\,}
\newcommand\ot{\otimes}
\newcommand\join{\vee}
\newcommand\meet{\wedge}
\newcommand\ttt{{\text{\rm t}}}
\newcommand\bfone{{\bf 1}}
\newcommand\aaa{\alpha}
\newcommand\bbb{\beta}
\newcommand\ccc{\gamma}
\newcommand\E{{\mathcal E}}
\newcommand\ghz{{\text{\rm GHZ}}}
\begin{document}

\preprint{JMP22-AR-00031}

\title{There exist infinitely many kinds of partial separability/entanglement}

\author{Kil-Chan Ha$^{1,*}$}
\author{Kyung Hoon Han$^2$}
\author{Seung-Hyeok Kye$^3$}
\affiliation{$^1$Faculty of Mathematics and Statistics, Sejong University, Seoul 143-747, Korea\\
$^2$Department of Data Science, The University of Suwon, Gyeonggi-do 445-743, Korea\\
$^3$Department of Mathematics and Institute of Mathematics, Seoul National University, Seoul 151-742, Korea}

\email[]{kcha@sejong.ac.kr,  kyunghoon.han@gmail.com,  kye@snu.ac.kr}

\date{\today}

\begin{abstract}
In tri-partite systems, there are three basic biseparability,
$A$-$BC$, $B$-$CA$ and $C$-$AB$ biseparability according to
bipartitions of local systems. We begin with three convex sets
consisting of these basic biseparable states in the three qubit
system, and consider arbitrary iterations of intersections and/or
convex hulls of them to get convex cones. One natural way to
classify tri-partite states is to consider those convex sets to
which they belong or do not belong. This is especially useful to
classify partial entanglement of mixed states. We show that the
lattice generated by those three basic convex sets with respect to
convex hull and intersection has infinitely many mutually distinct
members, to see that there are infinitely many kinds of three qubit
partial entanglement. To do this, we consider an increasing chain of
convex sets in the lattice and exhibit three qubit
Greenberger-Horne-Zeilinger diagonal states distinguishing those
convex sets in the chain.
\end{abstract}

\maketitle

\section{Introduction}

The notion of entanglement originated from quantum mechanics is now
widely recognized as one of the most important resources in the
current quantum information and computation theory. A quantum state
is called separable if it is a mixture of product states, and
entangled if it is not separable. In multi-partite systems, this
notion naturally depends on partitions of the local systems, and
classification of entanglement has been an important research topic
during a couple of last decades. See survey articles
\cite{{guhne_survey},{Horodecki-2009},{WGE}}. A multi-partite state
is called fully separable if it is separable in itself, and
partially separable if it is separable with respect to a partition
of the systems, or a mixture of them. A partially separable state
which is not fully separable is called partially entangled. A state
is also called genuinely entangled if it is not partially separable.

In case of three qubit system,
we call a state $A$-$BC$ biseparable if it is separable with respect to the bi-partition $A$-$BC$ of
systems, and similarly for other bi-partitions.
We will use the Greek letters for convex cones of basic biseparable states;
\begin{itemize}
\item
$\aaa$: the convex cone consisting of all unnormalized $A$-$BC$ biseparable states,
\item
$\bbb$: the convex cone consisting of all unnormalized $B$-$CA$ biseparable states,
\item
$\ccc$: the convex cone consisting of all unnormalized $C$-$AB$ biseparable states.
\end{itemize}
It is natural to classify three qubit states
according to the convex sets to which they belong, as it was suggested in \cite{dur-cirac-tarrach}.
In fact, the authors of \cite{dur-cirac-tarrach} classified three qubit states into the following five classes:
\begin{itemize}
\item
fully inseparable states; states which are not in $\aaa\cup\bbb\cup\ccc$,
\item
$1$-qubit biseparable states; states in $\aaa\setminus (\bbb\cup\ccc)$, $\bbb\setminus(\ccc\cup\aaa)$, $\ccc\setminus(\aaa\cup\bbb)$,
\item
$2$-qubit biseparable states; states in $(\aaa\cap\bbb)\setminus\ccc$, $(\bbb\cap\ccc)\setminus\aaa$, $(\ccc\cap\aaa)\setminus\bbb$,
\item
$3$-qubit biseparable states; states in $\aaa\cap\bbb\cap\ccc$ which are not fully separable,
\item
fully separable states.
\end{itemize}
Note that three qubit states belonging to one of the middle three classes are partially entangled.
We also note that it had been shown in \cite{bdmsst} that a three qubit state in the intersection of three convex sets need not to be
fully separable. Mixtures of $\aaa,\bbb$ and $\ccc$ also have been considered in \cite{acin}, and three qubit states in this class are
called biseparable states.
See \cite{{dur-cirac},{dur-vidal},{seevinck-uffink},{sz2011}}
for further discussion in this direction and extensions to arbitrary multi-partite cases.

The convex hull and intersection of two convex sets $\sigma_1$ and $\sigma_2$ will be denoted by
$\sigma_1\join\sigma_2$ and $\sigma_1\meet\sigma_2$,
respectively. With these notations, we list up the convex cones arising so far as follows:
\begin{equation}\label{ch1}
\aaa,\quad\bbb,\quad\ccc,
\quad
\aaa\meet\bbb,
\quad
\bbb\meet\ccc,
\quad
\ccc\meet\aaa,
\quad
\aaa\join\bbb\join\ccc,
\quad
\aaa\meet\bbb\meet\ccc.
\end{equation}
The authors of \cite{sz2012} considered the convex hulls of two basic biseparable states, and classified tri-partite states
by considering their intersections and complements to get $18$ kinds of partial separability(PS) classes. In this way,
the above list has been expanded to include the following convex cones;
\begin{equation}\label{ch2}
\aaa\join\beta,\qquad
\aaa\meet(\bbb\join\ccc),\qquad
(\aaa\join\bbb)\meet(\aaa\join\ccc),\qquad
(\aaa\join\bbb)\meet(\bbb\join\ccc)\meet(\ccc\join\aaa),
\end{equation}
together with those obtained by permutations of $\aaa$, $\bbb$ and  $\ccc$.
See also \cite{{han_kye_bisep_exam}} for examples distinguishing them, and \cite{{sz2015},{sz2018},{Szalay-2019}} for multi-partite cases.
Motivated by the list, the authors in \cite{han_kye_szalay}
considered the lattice ${\mathcal L}$ generated by $\aaa,\bbb$ and $\ccc$ with respect to the operations
$\join$ and $\meet$,
and showed that this lattice  violates the distributive rules; it satisfies the strict inclusion
\begin{equation}\label{diff}
\aaa\join(\bbb\meet\ccc)\lneqq
(\aaa\join\bbb)\meet(\aaa\join\ccc).
\end{equation}
Note that elements of the lattice ${\mathcal L}$ are convex cones in the $64$-dimensional affine space
of all self-adjoint three qubit matrices.
See \cite{{birkhoff},{freese},{salii}} for general theory of lattices.

Extending the classification in \cite{sz2012},
it is natural to say that three qubit states $\varrho_1$ and $\varrho_2$ are in the same
{\sl partial separability(PS) class} when the relation
$\varrho_1\in\sigma \Longleftrightarrow \varrho_2\in\sigma$
holds for every convex cone $\sigma\in{\mathcal L}$.
Suppose that a completely positive map $\Phi$ has a Kraus decomposition \cite{kraus} of the form
$$
\Phi(X)=\sum_j (V_{1j}\ot V_{2j}\ot V_{3j}) X (V_{1j}^*\ot
V_{2j}^*\ot V_{3j}^*),\qquad X\in M_2\ot M_2\ot M_2
$$
with $2\times 2$ matrices $V_{ij}$. Then it is clear that if a state $\varrho$ belongs to a convex cone $\sigma\in{\mathcal L}$
then $\Phi(\varrho)$ also belongs to the same convex cone $\sigma\in{\mathcal L}$.
Because SLOCC relation is described by such a transform \cite{bprst},
we see that PS classes give rise to a coarse-grained classification than SLOCC. In fact,
there exist only five PS classes for three qubit pure states; genuinely entangled states,
partially entangled states belonging to $\aaa$, $\bbb$, $\ccc$,
and fully separable states.  On the other hand, there are exactly six SLOCC classes; genuine entanglement
splits into two subclasses; GHZ states and W states \cite{dur-vidal}.

The purpose of this paper is to show that there exist infinitely many PS classes for mixed three qubit states.
To do this, we follow \cite[Theorem 1.23]{freese} to define the convex cones $f(\sigma)$ by
$$
f(\sigma)=\aaa\join(\bbb\meet(\ccc\join(\aaa\meet(\bbb\join(\ccc\meet\sigma)))))
$$
for  $\sigma\in{\mathcal L}$. We have
the increasing chain $\aaa \le f(\aaa)\le f^2(\aaa)\le\dots$
by the relation $\alpha\le f(\alpha)$. We will show that the strict inclusions
\begin{equation}\label{chain}
\aaa \lneqq f(\aaa)\lneqq\cdots \lneqq f^{n-1}(\aaa)\lneqq f^{n}(\aaa)\lneqq \cdots
\end{equation}
hold, by exhibiting three qubit Greenberger-Horne-Zeilinger diagonal states
$$
\varrho_n\in f^n(\aaa)\setminus f^{n-1}(\aaa),\qquad n=1,2,\dots.
$$
In the next section, we collect several known results to exhibit such states, and finally construct required states
in the Section 3.

\section{Greenberger-Horne-Zeilinger diagonal states}

Our main tool is the duality between matrices. We define the bilinear pairing $\lan a,b\ran=\tr(ab^\ttt)$ for two matrices $a$ and $b$.
Recall that $a$ is positive (semi-definite) if and only if $\lan a,b\ran\ge 0$ for every positive  $b$.
For a subset $C$ of self-adjoint matrices, we define the dual cone $C^\circ$ by
the set of all self-adjoint matrices $b$ satisfying $\lan a,b\ran\ge 0$ for every $a \in C$.
Then $C^{\circ\circ}$ is the smallest closed convex cone
containing $C$. In particular, we have $C^{\circ\circ}=C$ for a closed convex cone $C$. Therefore, we have
$a\in C$ if and only if $\lan a,b\ran\ge 0$ for every $b\in C^\circ$.
If we consider the convex cone of all separable unnormalized states then we see that a state $\varrho$ is entangled if and only if
there exists a self-adjoint $W$ in the dual cone such that $\lan W,\varrho\ran <0$. Such $W$ is called a witness \cite{terhal}.
Through the Choi isomorphism \cite{{choi75-10},{dePillis}},
witnesses can be interpreted \cite{{han_kye_tri},{kye-dual_multi}}
as multi-linear maps with $(n-1)$ variables when there are $n$ local systems.

In order to deal with GHZ diagonal states, we will restrict ourselves to the so called {\sf X}-shaped matrices whose entries are zero except
for diagonal and anti-diagonal ones. Therefore, a self-adjoint three qubit {\sf X}-shaped matrix is of the form
$$
X(a,b,c)= \left(
\begin{matrix}
a_1 &&&&&&& c_1\\
& a_2 &&&&& c_2 & \\
&& a_3 &&& c_3 &&\\
&&& a_4&c_4 &&&\\
&&& \bar c_4& b_4&&&\\
&& \bar c_3 &&& b_3 &&\\
& \bar c_2 &&&&& b_2 &\\
\bar c_1 &&&&&&& b_1
\end{matrix}
\right),
$$
for vectors
$a=(a_1,a_2,a_3,a_4),b=(b_1,b_2,b_3,b_4)\in\mathbb R^4$ and
$c=(c_1,c_2,c_3,c_4)\in\mathbb C^4$.
Complete characterization for {\sf X}-states to belong to convex cones appearing in (\ref{ch1}) and (\ref{ch2})
and their dual cones
have been found in \cite{{guhne10},{han_kye_optimal},{han_kye_bisep_exam},{han_kye_pe},{han_kye_lat},{han_kye_szalay},{Rafsanjani}}.
Important examples of {\sf X}-states are GHZ diagonal states.

The three qubit GHZ states \cite{{GHSZ},{GHZ}} are  given by
$$
|\ghz_{ijk}\ran=\frac 1{\sqrt 2}\left( |i\ran \ot|j\ran \ot|k\ran
+(-1)^i|\bar i\ran  \ot|\bar j \ran \ot|\bar k\ran \right)
\in \mathbb C^2\otimes\mathbb C^2\otimes\mathbb C^2,
$$
where $i+\bar i=1\mod 2$ and the index $ijk$ runs through $i,j,k\in\{0,1\}$.
We endow the indices with the lexicographic order to get eight vectors
$|\ghz_1\ran,|\ghz_2\ran,\dots, |\ghz_8\ran$.
A GHZ diagonal state is of the form
$$
\varrho_p=\sum_{i=1}^8 p_i|\ghz_i\ran\lan \ghz_i|
$$
for a probability distribution $p$. Then we see that $\varrho_p$ is an {\sf X}-state $X(a,b,c)$ with
$$
\begin{aligned}
a=b=&\frac 12 (p_1+p_8, p_2+p_7, p_3+p_6, p_4+p_5),\\
c=&\frac 12(p_1-p_8, p_2-p_7, p_3-p_6,p_4-p_5).
\end{aligned}
$$
In matrix form, we have
$$
\varrho_p
=\frac 12\left(
\begin{matrix}
p_1+p_8 &&&&&&& p_1-p_8\\
& p_2+p_7 &&&&& p_2-p_7 & \\
&& p_3+p_6 &&& p_3-p_6 &&\\
&&& p_4+p_5&p_4-p_5 &&&\\
&&&  p_4-p_5& p_4+p_5&&&\\
&&  p_3-p_6 &&& p_3+p_6 &&\\
&  p_2-p_7 &&&&& p_2+p_7 &\\
 p_1-p_8 &&&&&&& p_1+p_8
\end{matrix}
\right).
$$

Therefore, an {\sf X}-shaped state is GHZ diagonal if and only if $a=b$ and $c\in\mathbb R^4$.
We say that an {\sf X}-shaped self-adjoint matrix $X(a,b,c)$ is {\sl GHZ diagonal} when these conditions hold.
We denote by $V$ the eight dimensional real vector space consisting of all three qubit GHZ diagonal matrices.
For $W_q=\sum_{i=1}^8{q_i}|\ghz_i\ran\lan\ghz_i|$ with $q\in\mathbb R^8$, we also have
$$
\lan W_q,\varrho_p\ran=\sum_{i=1}^8p_iq_i=:\lan p,q\ran.
$$
From now on, we will identity the vector space $V$ with $\mathbb R^8$ through $\varrho_p\leftrightarrow p$
and $W_q\leftrightarrow q$. Then the convex cone
$$
\sigma_\ghz:=\sigma\cap V,\qquad \sigma\in{\mathcal L}
$$
can be realized as a convex cone in $\mathbb R^8$ through
the identification $V=\mathbb R^8$. For an example, the convex set of all three qubit GHZ diagonal states is
identified as the seven dimensional
simplex with the eight vertices given by the usual orthonormal basis $\{e_1,e_2,\dots, e_8\}$ of $\mathbb R^8$.

We denote by $\aaa_\xx$, $\bbb_\xx$ and $\ccc_\xx$ the convex cones of {\sf X}-shaped states belonging to $\aaa$, $\bbb$ and $\ccc$,
respectively. The extreme rays of them and their dual have been found in \cite[Section 3]{han_kye_pe}.
For an arbitrary convex cone $\sigma$ in the lattice ${\mathcal L}$, it was also shown in \cite{han_kye_lat} that
a GHZ diagonal state $\varrho$ belongs to $\sigma$ if and only if $\lan W,\varrho\ran\ge 0$ for every
$W\in\sigma^\circ$ which is GHZ diagonal. Furthermore, extreme rays of the convex cone $\aaa_\ghz$
are extreme rays of $\aaa_\xx$ which are GHZ diagonal. The same holds for $\bbb_\ghz$ and $\ccc_\ghz$.
We list them under the identification with $\mathbb R^8$.
All of them have common extreme rays
$$
\Delta=\{e_1+e_8,\ e_2+e_7,\ e_3+e_6,\ e_4+e_5\}
$$
in $\mathbb R^8$ which make diagonal states. For a given convex cone $C$, we denote by ${\mathcal E}(C)$ the
set of vectors which generate extreme rays of $C$. Then we have
$$
\begin{aligned}
\E(\aaa_\ghz)&=\Delta\cup
\{e_1+e_4,\
e_4+e_8,\
e_8+e_5,\
e_5+e_1,\
e_2+e_3,\
e_3+e_7,\
e_7+e_6,\
e_6+e_2\}\\
\E(\bbb_\ghz)&=\Delta\cup
\{e_1+e_3,\
e_3+e_8,\
e_8+e_6,\
e_6+e_1,\
e_2+e_4,\
e_4+e_7,\
e_7+e_5,\
e_5+e_2\},\\
\E(\ccc_\ghz)&=\Delta\cup
\{e_1+e_2,\
e_2+e_8,\
e_8+e_7,\
e_7+e_1,\
e_3+e_4,\
e_4+e_6,\
e_6+e_5,\
e_5+e_3\}.
\end{aligned}
$$

The convex cones $\aaa_\ghz^\circ$, $\bbb_\ghz^\circ$ and $\ccc_\ghz^\circ$ have common extreme rays
$$
{\sf E}=\{e_1,\ e_2,\ e_3,\ e_4,\ e_5,\ e_6,\ e_7,\ e_8\}
$$
in $\mathbb R^8$, and we also have
$$
\begin{aligned}
\E(\aaa_\ghz^\circ)
&={\sf E}\cup \{
-e_1+e_4+e_5+e_8,\
e_1-e_4+e_5+e_8,\
e_1+e_4-e_5+e_8,\
e_1+e_4+e_5-e_8,\\
&\phantom{\Delta^\circ\cup \{}
-e_2+e_3+e_6+e_7,\
e_2-e_3+e_6+e_7,\
e_2+e_3-e_6+e_7,\
e_2+e_3+e_6-e_7\}\\
\E(\bbb_\ghz^\circ)
&={\sf E}\cup \{
-e_1+e_3+e_6+e_8,\
e_1-e_3+e_6+e_8,\
e_1+e_3-e_6+e_8,\
e_1+e_3+e_6-e_8,\\
&\phantom{\Delta^\circ\cup \{}
-e_2+e_4+e_5+e_7,\
e_2-e_4+e_5+e_7,\
e_2+e_4-e_5+e_7,\
e_2+e_4+e_5-e_7\}\\
\E(\ccc_\ghz^\circ)
&={\sf E}\cup \{
-e_1+e_2+e_7+e_8,\
e_1-e_2+e_7+e_8,\
e_1+e_2-e_7+e_8,\
e_1+e_2+e_7-e_8,\\
&\phantom{\Delta^\circ\cup \{}
-e_3+e_4+e_5+e_6,\
e_3-e_4+e_5+e_6,\
e_3+e_4-e_5+e_6,\
e_3+e_4+e_5-e_6\}.
\end{aligned}
$$
A GHZ diagonal state $\varrho_p$ belongs to $\aaa$ if and only if $\lan p,q\ran\ge 0$ for every $q\in \E(\aaa_\ghz^\circ)$,
and similarly for $\bbb$ and $\ccc$.
It follows that
\begin{equation}\label{alpha}
\varrho_p \in \aaa_\ghz \Longleftrightarrow p_i \le p_j+p_k+p_\ell ~\text{for}~ \{i,j,k,\ell\}=\{1,4,5,8\}, \{2,3,6,7\},
\end{equation}
\begin{equation}\label{beta}
\varrho_p \in \bbb_\ghz \Longleftrightarrow p_i \le p_j+p_k+p_\ell ~\text{for}~ \{i,j,k,\ell\}=\{1,3,6,8\}, \{2,4,5,7\},
\end{equation}
\begin{equation}\label{gamma}
\varrho_p \in \ccc_\ghz \Longleftrightarrow p_i \le p_j+p_k+p_\ell ~\text{for}~ \{i,j,k,\ell\}=\{1,2,7,8\}, \{3,4,5,6\}.
\end{equation}
Therefore, the convex cone $\alpha_\ghz$ in $\mathbb R^8$ is determined by finitely many homogeneous linear equations,
and so it is a polytope. The same is true for $\sigma_\ghz$ for every $\sigma\in{\mathcal L}$.
This is also the case for arbitrary multi-qubit systems. Those polytopes consisting of multi-qubit
biseparable and fully biseparable states have been explored recently \cite{han_kye_poly}.

\section{Infinitely many convex cones in the lattice}

We denote by $\sigma_0, \sigma_1, \sigma_2,\dots$ the convex cones in the chain (\ref{chain}), that is, we define
$$
\sigma_0=\aaa,\qquad \sigma_n=f(\sigma_{n-1}),\qquad n=1,2,\dots.
$$
We will exhibit a GHZ diagonal state $\varrho_n\in \sigma_n\setminus \sigma_{n-1}$ for each $n=1,2,\dots$ to show that the lattice
contains infinitely {\color{red}many} mutually distinct convex cones. To do this, we define
$$
\sigma_n^1=\ccc\meet \sigma_n,\quad
\sigma_n^2=\bbb\join \sigma_n^1,\quad
\sigma_n^3=\aaa\meet \sigma_n^2,\quad
\sigma_n^4=\ccc\join \sigma_n^3,\quad
\sigma_n^5=\bbb\meet \sigma_n^4,\quad
$$
for $n=0,1,2,\dots$. Then we have
$$
\sigma_{n+1}=f(\sigma_n)=\aaa\join(\bbb\meet(\ccc\join(\aaa\meet(\bbb\join(\ccc\meet\sigma_n)))))=\aaa\join \sigma_n^5.
$$

We define the states $\varrho_n^\bbb$ by
$$
\begin{aligned}
\varrho_n^\bbb
&=\begin{cases}
(2n+2)(e_1+e_3)+(e_4+e_5),\quad &n\ {\text{\rm is even}}\\
(2n+1)(e_2+e_5)+(e_2+e_4),\quad &n\ {\text{\rm is odd}}\end{cases}
\end{aligned}
$$
for $n=0,1,2,\dots$. Then we have $\varrho_n^\bbb\in \bbb_\ghz$ since it is the sum of states in $\E(\bbb_\ghz)$. We also define
the states $\varrho_n^\ccc\in\ccc_\ghz$ and $\varrho_n^\aaa\in\aaa_\ghz$ by
$$
\begin{aligned}
\varrho_n^\ccc
&=\begin{cases}
(2n+2)(e_3+e_5),\quad &n\ {\text{\rm is even}}\\
(2n+2)(e_1+e_2),\quad &n\ {\text{\rm is odd}}\end{cases}\\
\varrho_n^\aaa
&=\begin{cases}
(2n+4)(e_2+e_3)+(e_4+e_5),\quad &n\ {\text{\rm is even}}\\
(2n+3)(e_1+e_5)+(e_1+e_4),\quad &n\ {\text{\rm is odd}}\end{cases}
\end{aligned}
$$

Now, we define inductively the sequence of states by
$$
\varrho_0=(2,1,0,1,1,0,1,0),\qquad \varrho_{n+1}=\varrho_n+\varrho_n^\bbb +\varrho_n^\ccc+\varrho_n^\aaa,$$
where
$$
\varrho_n^\bbb +\varrho_n^\ccc+\varrho_n^\aaa =
\begin{cases}(2n+2,2n+4,6n+8,2,2n+4,0,0,0),\qquad n\ {\text{\rm is even}} \\ (4n+6,4n+4,0,2,4n+4,0,0,0),\qquad n\ {\text{\rm is odd}} \end{cases}
$$
for $n=0,1,2,\dots$. Then $\varrho_0=(2,1,0,1,1,0,1,0)$ belongs to the convex cone $\sigma_0=\aaa$ by $(\ref{alpha})$ or the decomposition $\varrho_0=(e_1+e_4)+(e_1+e_5)+(e_2+e_7)$.
When $n$ is even, we directly compute
$$
\begin{aligned}
\varrho_{n}=&\varrho_0+\sum_{k=0}^{n-1}\bigl(\varrho_k^\bbb +\varrho_k^\ccc+\varrho_k^\aaa\bigr)\\
=&\varrho_0+\sum_{k=0}^{n/2-1}\bigl(\varrho_{2k}^\bbb +\varrho_{2k}^\ccc+\varrho_{2k}^\aaa\bigr)+
\sum_{k=0}^{n/2-1}\bigl(\varrho_{2k+1}^\bbb +\varrho_{2k+1}^\ccc+\varrho_{2k+1}^\aaa\bigr)\\
=&(2,1,0,1,1,0,1,0)+\sum_{k=0}^{n/2-1}(4k+2,4k+4,12k+8,2,4k+4,0,0,0)+\sum_{k=0}^{n/2-1}(8k+10,8k+8,0,2,8k+8,0,0,0)\\
=&(2,1,0,1,1,0,1,0)+\sum_{k=0}^{n/2-1}(12k+12,12k+12,12k+8,4,12k+12,0,0,0)\\
=&\left({3\over2}n^2+3n+2,{3\over2}n^2+3n+1,{3\over2}n^2+n,2n+1,{3\over2}n^2+3n+1,0,1,0\right)
\end{aligned}
$$
When $n$ is odd, we apply this formula to $\varrho_{n-1}$ to obtain
$$
\begin{aligned}
\varrho_{n}=&\varrho_{n-1}+\varrho_{n-1}^\bbb +\varrho_{n-1}^\ccc+\varrho_{n-1}^\aaa\\
=&\varrho_{n-1}+(2n,2n+2,6n+2,2,2n+2,0,0,0)\\
=&\left({3\over2}n^2+2n+{1\over2},{3\over2}n^2+2n+{3\over2},{3\over2}n^2+4n+{5\over2},2n+1,{3\over2}n^2+2n+{3\over2},0,1,0\right)
\end{aligned}
$$
For brevity, we can write
$$
\varrho_n=
\begin{cases}
(a_n+1, a_n, a_n-2n-1, 2n+1, a_n, 0,1,0),\qquad n\ {\text{\rm is even}}\\
(a_n-1, a_n, a_n+2n+1, 2n+1, a_n, 0,1,0),\qquad n\ {\text{\rm is odd}}
\end{cases}
$$
with
$$
a_n=\begin{cases}\frac 32n^2+3n+1,\quad &n\ {\text{\rm is even}}\\
\frac 32n^2+2n+\frac 32,\quad &n\ {\text{\rm is odd.}}\end{cases}
$$

Suppose that
$\varrho_n\in \sigma_n$. By (\ref{gamma}), $\varrho_n$ belongs to $\ccc$, and so we have
$\varrho_n\in \sigma_n^1=\ccc\meet \sigma_n$, and $\varrho_n+\varrho_n^\bbb\in \sigma_n^2=\bbb\join \sigma_n^1$.
 We see that
$$ 
\varrho_n+\varrho_n^\bbb =
\begin{cases}
(a_n+2n+3, a_n, a_n+1, 2n+2, a_n+1, 0,1,0),\qquad n\ {\text{\rm is even}}\\
(a_n-1, a_n+2n+2, a_n+2n+1, 2n+2, a_n+2n+1, 0,1,0),\qquad n\ {\text{\rm is odd}}
\end{cases}
$$
belongs to $\aaa$ by (\ref{alpha}), and so we have $\varrho_n+\varrho_n^\bbb\in \sigma_n^3 = \aaa\meet \sigma_n^2$ and $\varrho_n+\varrho_n^\bbb+\varrho_n^\ccc \in \sigma_n^4 = \ccc\join \sigma_n^3$. Now, we apply (\ref{beta}) to see that
$$
\varrho_n+\varrho_n^\bbb+\varrho_n^\ccc =
\begin{cases}
(a_n+2n+3, a_n, a_n+2n+3, 2n+2, a_n+2n+3, 0,1,0),\qquad n\ {\text{\rm is even}}\\
(a_n+2n+1, a_n+4n+4, a_n+2n+1, 2n+2, a_n+2n+1, 0,1,0),\qquad n\ {\text{\rm is odd}}
\end{cases}
$$
belongs to $\bbb$. Therefore, we have $\varrho_n+\varrho_n^\bbb+\varrho_n^\ccc \in \bbb\meet\sigma_n^4$
and $\varrho_{n+1}=\varrho_n+\varrho_n^\bbb+\varrho_n^\ccc+\varrho_n^\aaa \in \sigma_{n+1}=\aaa\join\sigma_n^5$.
By induction, we conclude that $\varrho_n\in \sigma_n$ for each $n=0,1,2,\dots$.

In order to show that $\varrho_{n+1}\notin \sigma_{n}$, we look for witnesses $W_n\in \sigma_n^\circ$
satisfying the relation $\lan W_n, \varrho_{n+1}\ran<0$. We define
$$
W_n=\begin{cases}
(0,1,-1,0,0,2n+1,2n+1,2n),\quad &n\ {\text{\rm is even}}\\
(0,-1,1,0,0,2n+1,2n+1,2n),\quad &n\ {\text{\rm is odd}}
\end{cases}
$$
We first show that $W_n\in \sigma_n^\circ$. First of all, $W_0$ generates an extreme ray of $\aaa_\ghz^\circ$,
and so belongs to $\sigma_0^\circ$.
Suppose that $W_n\in \sigma_n^\circ$. When $n$ is even, we add $e_1-e_2+e_7+e_8\in\ccc^\circ$ to get
$$
(1,0,-1,0,0,2n+1,2n+2,2n+1)\in \ccc^\circ\join \sigma_n^\circ=(\ccc\meet \sigma_n)^\circ=(\sigma_n^1)^\circ.
$$
Since it is the sum of $(e_1-e_3+e_6+e_8)$ and the nonnegative multiples of $e_i$, it belongs to $\bbb^\circ\meet (\sigma_n^1)^\circ=(\bbb\join \sigma_n^1)^\circ=(\sigma_n^2)^\circ$.
Now, we add $-e_2+e_3+e_6+e_7\in \aaa^\circ$ to get
$$
(1,-1,0,0,0,2n+2,2n+3,2n+1)\in \aaa^\circ\join (\sigma_n^2)^\circ=(\aaa\meet \sigma_n^2)^\circ=(\sigma_n^3)^\circ.
$$
It is the sum of $(e_1-e_2+e_7+e_8)$ and the nonnegative multiples of $e_i$, and so it belongs to $\ccc^\circ\meet (\sigma_n^3)^\circ=(\ccc\join \sigma_n^3)^\circ=(\sigma_n^4)^\circ$.
Finally, we add $-e_1+e_3+e_6+e_8\in\bbb^\circ$ to get
$$
W_{n+1}=(0,-1,1,0,0,2n+3,2n+3,2n+2)\in \bbb^\circ\join (\sigma_n^4)^\circ=(\bbb\meet \sigma_n^4)^\circ=(\sigma_n^5)^\circ.
$$
Since it is the sum of $(-e_2+e_3+e_6+e_7)$ and the nonnegative multiples of $e_i$, it belongs to 
$\bbb^\circ\meet (\sigma_n^5)^\circ=(\bbb\join \sigma_n^5)^\circ=(\sigma_{n+1})^\circ$.
When $n$ is odd, we add up
$-e_1+e_2+e_7+e_8\in\ccc^\circ$,
$e_2-e_3+e_6+e_7\in \aaa^\circ$ and
$e_1-e_3+e_6+e_8\in\bbb^\circ$
in the above argument, to see $W_{n+1}\in \sigma_{n+1}^\circ$.
By induction, we conclude that $W_n\in \sigma_n^\circ$ for each $n=0,1,2,\dots.$

Finally, we show $\lan W_n, \varrho_{n+1}\ran<0$. If $n$ is even then we have
$$
\lan W_n, \varrho_{n+1}\ran
=a_{n+1}-\bigl(a_{n+1}+2(n+1)+1\bigr)+(2n+1)=-2<0.
$$
Similarly, we also have $\lan W_n,\varrho_{n+1}\ran=-2$ for odd $n$.

\section{Conclusion}

In this paper, we have shown that there exist infinitely many PS classes for mixed three qubit states.
All of them have different properties with respect to partial separability. For examples,
states in the difference between two convex sets in (\ref{diff}) have the following common properties;
\begin{itemize}
\item
they are mixtures of $A$-$BC$ biseparable states and $B$-$CA$ biseparable states;
\item
they are mixtures of $A$-$BC$ biseparable states and $C$-$AB$ biseparable states;
\item
nevertheless, they are not a mixture of $A$-$BC$ biseparable states and simultaneously $B$-$CA$ and $C$-$AB$ biseparable states.
\end{itemize}
Because the relation
$$
\aaa\join(\bbb\meet\ccc)\le
\aaa\join(\bbb\meet(\ccc\join(\aaa\meet\sigma)))\le
\aaa\join(\bbb\meet(\ccc\join\aaa))\le
(\aaa\join\bbb)\meet(\aaa\join\ccc)
$$
holds for every $\sigma\in{\mathcal L}$, we see that all the states $\varrho_n$ also
belong to the difference between two convex cones in (\ref{diff}).
Therefore, they retain mutually different partial separability properties
even though they share the above three properties.

One may hope to find a canonical way to list up all the convex cones in the lattice ${\mathcal L}$.
To do this, we have to find all the possible lattice identities among the generators $\aaa$, $\bbb$ and $\ccc$.
The first candidate is the following
$$
(\aaa\meet\bbb)\join(\aaa\meet\ccc)
=\aaa\meet [\aaa\join(\bbb\meet\ccc)]\meet[\bbb\join(\ccc\meet\aaa)]\meet[\ccc\join(\aaa\meet\bbb)]
$$
together with the identities by permutation and lattice dual. It was shown in \cite{han_kye_lat} that
these identities hold among {\sf X}-states. It would be nice to know
if they hold for arbitrary three qubit states.
In this regards, it is natural to ask if the lattice ${\mathcal L}$ is free or not.

 If two convex sets $\sigma_1$ and $\sigma_2$ in the lattice satisfy $\sigma_1\subsetneqq\sigma_2$
then states in $\sigma_2\setminus \sigma_1$
may be considered to contain more entanglement than those in $\sigma_1$.
It would be nice to find a new kind of entanglement measure to confirm this claim.
For a given two partially separable states, it is very hard to determine if they are in the same PS class or not.
Finding invariants to distinguish PS classes might be a challenging project. Such invariants have been found
\cite{sz2012} for those PS classes listed in (\ref{ch1}) and (\ref{ch2}).

The result of the present paper can be generalized to arbitrary multi-partite systems $M_{A_1}\ot M_{A_2}\otimes \cdots\otimes M_{A_n}$
by the same argument as in \cite[Section 4]{han_kye_szalay}.
For a given partition $\Pi$ of local systems $\{A_1,A_2,\dots,A_n\}$, we denote by $\aaa_\Pi$
the convex cone consisting of all partially separable (unnormalized) states with respect to the partition $\Pi$.
When $M_{A_i}$ is the $d_i\times d_i$ matrices, we denote by ${\mathcal L}_{d_1,d_2,\dots,d_n}$
the lattice generated by $\alpha_\Pi$ through all nontrivial partitions $\Pi$
with respect to two operations, intersection and convex hull.
So far, we have shown that the lattice ${\mathcal L}_{2,2,2}$ is infinite.
If we apply the argument of \cite[Section 4]{han_kye_szalay} on
$\varrho \in \left(\alpha \meet (\beta \join \gamma)\right) ~\backslash~ (\alpha \meet \beta) \join (\alpha \meet \gamma)$
to $\varrho_n \in f^n(\aaa)\setminus f^{n-1}(\aaa)$, then we see that the lattice ${\mathcal L}_{d_1,d_2,\dots,d_n}$ is also infinite.

\begin{acknowledgments}
The authors are grateful to Szil\'ard Szalay for valuable comments on the manuscript.

Both Kyung Hoon Han and Seung-Hyeok Kye were partially supported by the National Research Foundation of Korea (Grant No. NRF-2020R1A2C1A01004587) and Kil-Chan Ha was partially supported by the National Research Foundation of Korea (Grant No. NRF-2016R1D1A1A09916730).
\end{acknowledgments}

\section*{Data Availability}
The data that supports the findings of this study are available within the article.


\section*{References}
\begin{thebibliography}{99}
\bibitem{acin}
A. Acin, D. Bru\ss, M. Lewenstein and A. Sanpera, \it Classification
of mixed three-qubit states, \rm Phys. Rev. Lett. {\bf 87} (2001),
040401.

\bibitem{bdmsst}
C. H. Bennett, D. P. DiVincenzo, T. Mor, P. W. Shor, J. A. Smolin
and B. M. Terhal, \it Unextendible product bases and bound
entanglement, \rm Phys. Rev. Lett. \bf 82 \rm (1999), 5385--5388.

\bibitem{bprst}
C. H. Bennett, S. Popescu, D.  Rohrlich, J. A. Smolin and A. V. Thapliyal,
\it Exact and asymptotic measures of multipartite pure-state entanglement,
\rm Phys. Rev. A {\bf 63} (2001), 012307.


\bibitem{birkhoff}
G. Birkhoff, Lattice Theory, 3rd ed., Amer. Math. Soc. Coll. Publ.
Vol XXV, Amer. Math. Soc. 1967


\bibitem{choi75-10}
M.-D. Choi, \it Completely positive linear maps on complex matrices,
\rm Linear Alg. Appl. \bf 10 \rm (1975), 285--290.


\bibitem{dePillis}
J. de Pillis,
\it Linear transformations which preserve Hermitian and positive semidefinite operators,
\rm Pacific J. Math. {\bf 23} (1967), 129--137.
\bibitem{dur-cirac}
W. D\" ur and J. I. Cirac, \it Classification of multi-qubit mixed
states: separability and distillability properties, \rm Phys. Rev. A
{\bf 61} (2000), 042314.

\bibitem{dur-cirac-tarrach}
W. D\" ur, J. I. Cirac and R. Tarrach, \it Separability and
Distillability of Multiparticle Quantum Systems, \rm Phys. Rev.
Lett. {\bf 83} (1999), 3562--3565.

\bibitem{dur-vidal}
W. D\" ur, G. Vidal and J. I. Cirac, \it Three qubits can be
entangled in two inequivalent ways, \rm Phys. Rev. A {\bf 62}
(2000), 062314.

\bibitem{freese}
R. Freese, J. Jezek and J. B. Nation, Free Lattice, Math. Surv
Monog. Vol 42, Amer. Math. Soc., 1995.



\bibitem{GHSZ}
D. M. Greenberger, M. A. Horne, A. Shimony and A. Zeilinger,
\it Bell's theorem without inequality,
\rm Am. J. Phys. {\bf 58} (1990), 1131--1143.


\bibitem{GHZ}
D. M. Greenberger, M. A. Horne and A. Zeilinger,
\it Going beyond Bell's theorem,
\rm in Bell’s Theorem, Quantum Theory and Conceptions of the Universe,
Fundamental Theories of Physics, Vol. 37 (1989), 73--76.

\bibitem{guhne10}
O. G\" uhne and M. Seevinck, \it Separability criteria for genuine
multiparticle entanglement, \rm New J. Phys. {\bfone 2} (2010),
053002.

\bibitem{guhne_survey}
O. G\" uhne and G. Toth,
\it Entanglement detection,
\rm Phys. Rep. \bf 474 \rm (2009), 1--75.


\bibitem{han_kye_tri}
K. H. Han and S.-H, Kye, \it Various notions of positivity for
bi-linear maps and applications to tri-partite entanglement, \rm J.
Math. Phys. {\bf 57} (2016), 015205.

\bibitem{han_kye_optimal}
K. H. Han and S.-H, Kye, \it Construction of multi-qubit optimal
genuine entanglement witnesses, \rm J. Phys. A: Math. Theor. {\bf
49} (2016), 175303.

\bibitem{han_kye_bisep_exam}
K. H. Han and S.-H. Kye, \it Construction of three-qubit biseparable
states distinguishing kinds of entanglement in a partial
separability classification, \rm Phys. Rev. A, {\rm 99} (2019),
032304.

\bibitem{han_kye_pe}
K. H. Han and S.-H, Kye, \it On the convex cones arising from
classifications of partial entanglement in the three qubit system,
\rm J. Phys. A: Math. Theor. {\bf 53} (2020), 015301.

\bibitem{han_kye_lat}
K. H. Han and S.-H. Kye,
\it Criteria for partial entanglement of three qubit states arising from distributive rules,
\rm Quantum Inf. Process. {\bf 20} (2021), 151.

\bibitem{han_kye_poly}
K. H. Han and S.-H. Kye,
\it Polytope structures for Greenberger-Horne-Zeilinger diagonal states,
\rm J. Phys. A: Math. Theor. {\bf 54} (2021), 455302.

\bibitem{han_kye_szalay}
K. H. Han, S.-H. Kye and S Szalay, \it Partial
separability/entanglement violates distributive rules, \rm Quantum
Inf. Process. {\bf 19} (2020), 202.

\bibitem{Horodecki-2009}
R. Horodecki, P. Horodecki, M. Horodecki and K. Horodecki, \it
Quantum entanglement, \rm Rev. Mod. Phys. {\bf 81} (2009), 865--942.


\bibitem{kraus}
K. Kraus,
\it Operations and effects in the Hilbert space
formulation of quantum theory,
\rm Foundations of quantum mechanics and ordered
linear spaces (Marburg, 1973), pp. 206--229. Lecture Notes
in Phys., Vol. 29,  Springer, 1974.

\bibitem{kye-dual_multi}
S.-H. Kye,
\it Three-qubit entanglement witnesses with the full spanning properties,
\rm J. Phys. A: Math. Theor. {\bf 48} (2015), 235303.

\bibitem{salii}
V. N. Salii, Lattices with unique complements, Trans. Math. Monog.
Vol 69, Amer. Math. Soc. 1988.

\bibitem{Rafsanjani}
S. M. H. Rafsanjani, M. Huber, C. J. Broadbent and J. H. Eberly \it
Genuinely multipartite concurrence of N-qubit X matrices, \rm Phys.
Rev. A {\bf 86} (2012), 062303.

\bibitem{seevinck-uffink}
M. Seevinck and J. Uffink, \it Partial separability and etanglement
criteria for multiqubit quantum states, \rm Phys. Rev. A {\bf 78}
(2008), 032101.


\bibitem{sz2011}
Sz. Szalay, \it Separability criteria for mixed three-qubit states,
\rm Phys. Rev. A {\bf 83} (2011), 062337.

\bibitem{sz2015}
Sz. Szalay, \it Multipartite entanglement measures, \rm Phys. Rev. A
{\bf 92} (2015), 042329.

\bibitem{sz2018}
Sz. Szalay, \it The classification of multipartite quantum
correlation, \rm J. Phys. A: Math. Theor. {\bf 51} (2018), 485302 .

\bibitem{Szalay-2019}
Sz. {\relax Sz}alay, \it $k$-stretchability of entanglement, and the
duality of $k$-separability and $k$-producibility, \rm Quantum {\bf
3} (2019), 204.

\bibitem{sz2012}
Sz. Szalay and Z. K\" ok\' enyesi, \it Partial separability
revisited: Necessary and sufficient criteria, \rm Phys. Rev. A {\bf
86}, 032341 (2012).

\bibitem{terhal}
B. M. Terhal, \it Bell Inequalities and the Separability Criterion,
\rm Phys. Lett. A \bf 271 \rm (2000), 319--326.

\bibitem{WGE}
M. Walter, D. Grossy and J. Eisert,
\it Multi-partite entanglement, \rm in
\lq\lq Quantum Information: From Foundations to Quantum Technology Applications, Volumes 1 and 2\rq\rq\ edited by
D. Bru\ss\ and G. Leuchs, Wiley-VCH Verlag, 2019.


\end{thebibliography}
\end{document}